\documentstyle[preprint,aps]{revtex}

\begin{document}                                                

\title{The Origin of Baryonic Matter in the Universe: A Brief Review}

\author{Marcelo Gleiser\footnote{NSF Presidential Faculty Fellow. Plenary
talk given at the Brazilian Meeting on Particles and Fields, Caxamb\'u, October
24-28, 1995.}}
\address{ Department of Physics and Astronomy, Dartmouth College,
Hanover, NH 03755 }
\date{ DART-HEP-96/02$~~~~~~~~~$January 1996}

\maketitle                       
\begin{abstract}
In this talk I briefly review the main ideas and challenges
involved in the
computation of the observed baryonic excess in the Universe. 
\end{abstract}

\section{Evidence for Baryonic Asymmetry}

One of the outstanding challenges of the interface between particle physics and
cosmology is the explanation for the observed baryonic asymmetry in the 
Universe \cite{KT}.
It is by now quite clear that there is indeed an excess of baryons over
antibaryons in the Universe. A strong constraint on the baryonic asymmetry
comes from big-bang nucleosynthesis, setting the net baryon number density 
($n_{{\rm B}}$) 
to photon entropy density ($s$) ratio at
about
$n_{{\rm B}}/s \equiv {{n_{{\rm b}}-n_{{\rm {\bar b}}}}\over s} \sim 8\times
10^{-11}$.
Within our solar system, there is no evidence
that antibaryons are primordial. Antiprotons found in cosmic rays at a
ratio of $N_{{\bar {\rm p}}}/N_{\rm p}\sim 10^{-4}$ are 
secondaries from collisions with the interstellar medium and do no indicate
the presence of primary antimatter within our galaxy \cite{STEIGMAN}. 

We
could imagine that in clusters of galaxies there would be antimatter
galaxies as 
well as galaxies. However, this being the case we should observe high energy
$\gamma$-rays from nucleons of galaxies annihilating with antinucleons of
``antigalaxies''. The fact that these are not detected rules out the presence of
both galaxies and antigalaxies on nearby
clusters, which typically have about $10^{14}~  M_{\odot}$ or so of material. 
For scales larger
than galactic clusters there is no observational evidence for the absence of
primordial antimatter. 

We could also imagine a baryon-symmetric 
Universe with large domains of
matter and antimatter separated over vast distances.
However, a simple cosmological argument rules out this possibility. In a 
locally baryon-symmetric Universe, nucleons remain in chemical equilibrium
with antinucleons down to temperatures of about $T\sim 22$ MeV or so, when
$n_{\rm b}/s 
\sim n_{{\bar {\rm b}}}/s \sim ~ 7\times 10^{-20}$. Annihilation is so efficient
as to become catastrophic! To avoid this annihilation, and still obey
the nucleosynthesis bound with a baryon-symmetric Universe, 
we need a mechanism to separate nucleons and
antinucleons by $T\sim 38$ MeV, when  $n_{\rm b}/s\sim n_{{\bar {\rm b}}}/s 
\sim ~ 8\times 10^{-11}$. However, at $T\sim 38$ MeV, the horizon contained
only about $10^{-7}M_{\odot}$, making separation of matter and antimatter
on scales of $10^{14}M_{\odot}$ causally impossible. It seems that we 
must settle for a primordial baryon asymmetry.

\section{ The Sakharov Conditions and GUT Baryogenesis}

Given that the evidence is for a Universe with a primordial baryon asymmetry,
we have two choices; either this asymmetry is the result of 
an initial condition, or
it was attained through dynamical processes that took place in the
early Universe. In 1967, just a couple of years after the discovery of
the microwave background radiation, 
Sakharov wrote a ground-breaking work in which he 
appealed to the drastic environment of the early stages of the hot big-bang
model to spell out the
3 conditions for dynamically generating the baryon asymmetry of the 
Universe \cite{SAKHAROV}. Here they are, with some modifications:

\noindent {\bf i)} Baryon number violating interactions: Clearly, if we are to
generate any excess baryons, our model must have interactions which violate
baryon number. However, the same interactions also produce antibaryons at
the same rate. We need a second
condition;

\noindent {\bf ii)} C and CP violating interactions: Combined violation of 
charge conjugation (C) and charge conjugation combined with parity (CP) 
can provide a bias to enhance the production of
baryons over antibaryons. However, in thermal equilibrium
$n_{{\rm b}}=n_{{\rm {\bar b}}}$, and  any asymmetry would be wiped out.
We need a third condition;

\noindent {\bf iii)} Departure from thermal equilibrium: Nonequilibrium
conditions guarantee that the phase-space density of baryons and antibaryons
will not be the same. Hence, provided there is no entropy production later
on, the net ratio $n_{{\rm B}}/s$ will remain constant. 

Given the above conditions, we have to search for the particle physics models
that both satisfy them and are capable of 
generating the correct asymmetry.
The first models that attempted to compute the baryon asymmetry
dynamically were Grand Unified Theory (GUT) models \cite{GUTBAR}. 
GUT models naturally satisfy conditions i) and 
ii); by construction, as strong and electroweak interactions are
unified, quarks and leptons appear as members of a common irreducible
representation of the GUT gauge group. Thus, gauge bosons mediate
interactions in which baryons can decay into leptons, leading to baryon
number violation. C and CP violation can be built into the models to
at least be consistent with the observed violation in the standard model. C
is maximally violated by weak interactions and CP violation is observed in the
neutral kaon system. 
One expects that C and CP violation
will be manifest in all sectors of the theory including the
superheavy boson sector ({\it e.g.},
$X\rightarrow q q$ with branching ratio $r$, 
and ${\bar X} \rightarrow {\bar q}{\bar q}$, with branching ratio ${\bar r}
 \neq r$).

Condition iii), departure from thermal equilibrium is provided by the
expansion of the Universe. In order for local thermal equilibrium to be
maintained in the background of an expanding Universe, the reactions that
create and destroy the heavy bosons $X$ and ${\bar X}$ (decay, annihilation, and
their inverse processes) must occur rapidly
with respect to the expansion rate of the Universe, $H={{{\dot R}}\over R}
\simeq T^2/M_{{\rm pl}}$, where $R(t)$ is the scale factor (the dot means time
derivative), $T$ is the 
temperature, and $M_{{\rm pl}}=1.2\times 10^{19}~$GeV is the Planck mass. A
typical mechanism of GUT baryogenesis is known as the ``out-of-equilibrium
decay scenario''; one insures that the heavy $X$ bosons have a 
long enough lifetime so that their
inverse decays go out of equilibrium as they are still
abundant. Baryon number is produced by the free decay of the heavy $X$s, as the
inverse rate is shut off.

Interesting as they are, GUT models of
baryogenesis have serious obstacles to overcome.
An obvious one is the lack of experimental confirmation for the main
prediction of GUTs, the decay of the proton. One can, however, build models
(invoking -or not- supersymmetry)
in which the lifetime surpases the limits of present experimental sensitivity.
A second obstacle is the production of magnetic monopoles predicted to happen 
as the GUT 
semi-simple
group is broken into subgroups that involve a $U(1)$. The existence of
such monopoles was one of the original motivations for inflationary models
of cosmology. As is well known, the existence of an inflationary, or
superluminal, expansion of the Universe will efficiently dilute any
unwanted relics from a GUT-scale transition (and before). Unfortunately,
inflation would also dilute badly wanted relics, such as the excess baryons
produced, say, by the out-of-equilibrium decay scenario mentioned above. One
way of bypassing this diluting effect is to have inflation followed by
efficient reheating to temperatures of about $10^{14}$ GeV, so  that
the processes responsible for baryogenesis could be reignited. Unfortunately,
reheating temperatures are usually much lower than this ($T_{{\rm reh}}<10^{12}
$ GeV, and $<10^9$ GeV for supersymmetric models due to nucleosynthesis
constraints on
gravitino decays), posing a serious problem for GUT baryogenesis. [Recent
work indicates that reheating temperatures could be much higher than previously
indicated, although it is too early to tell \cite{REHEAT}.] 

Finally,
a third obstacle to GUT baryogenesis comes from nonperturbative electroweak
processes. The vacuum manifold of the electroweak model exhibits a very
rich structure, with degenerate minima separated by energy barriers (in field
configuration space). Different minima have different baryon (and lepton) 
number, with
the net difference between two minima 
being given by the number of families. Thus, for the standard
model, each jump between two adjacent minima leads to the creation of 3
baryons and 3 leptons, with net $B-L$ conservation and $B+L$ violation.
At $T=0$, tunneling between adjacent minima is mediated
by instantons, and, as shown by 't Hooft \cite{tHOOFT}, the tunneling rate
is suppressed by the weak coupling constant ($\Gamma \sim e^{-4\pi/\alpha_{{\rm
W}}}\sim 10^{-170}$). That is why the proton is stable.
However, as pointed out by Kuzmin, Rubakov, and Shaposhnikov, at
finite temperatures ($T\sim 100$ GeV), one could hop over the barrier, 
tremendously enhancing the rate of baryon number violation \cite{KRS}. 
The height of
the barrier is given by the action of an unstable static solution of the field
equations known as the sphaleron \cite{SPHALERON}. 

Being a thermal process,
the rate of baryon number violation is controlled by the energy of the
sphaleron configuration, $\Gamma\sim {\rm exp}[-\beta E_{\rm S}]$, with
$E_{\rm S}\simeq M_{{\rm W}}/\alpha_{{\rm W}}$, where $M_{{\rm W}}$ is the 
W-boson mass. Note that $ M_{{\rm W}}/\alpha_{{\rm W}}= \langle\phi\rangle 
/g$, where $\langle\phi\rangle$ is 
the vacuum expectation value of the Higgs field. For temperatures above 
the critical temperature for electroweak symmetry restoration, it has been
shown that sphaleron processes are not exponentially suppressed, with the
rate being roughly $\Gamma\sim (\alpha_{{\rm W}}T)^4$ \cite{HIGHTSPH}.
Even though this
opens the possibility of generating the baryonic asymmetry at the electroweak
scale, it is bad news for GUT baryogenesis. Unless the original GUT
model was $B-L$ conserving, any net baryon number generated then would be
brought to zero by the efficient anomalous electroweak processes. There are
several alternative models for baryogenesis invoking more or less exotic
physics. The interested reader is directed to the review by Olive, listed
in Ref. 1. I now move
on to discuss the promises and challenges of electroweak baryogenesis.

\section{ Electroweak Baryogenesis}
     
As pointed out above, temperature effects can lead to efficient baryon
number violation at the electroweak scale. Can the  other two
Sakharov conditions
be satisfied in the early Universe so that the observed baryon number 
could be generated during the electroweak phase transition?
The short answer is that in principle yes, but probably 
not
in the context of the minimal standard model. Let us first see why 
it is possible to satisfy all conditions for baryogenesis in the context
of the standard model. 

Departure from thermal equilibrium is obtained by invoking a first order
phase transition. After summing over matter and gauge fields, one obtains
a temperature corrected effective potential for the magnitude of the Higgs
field, $\phi$. The potential describes two phases, the symmetric
phase with $\langle\phi\rangle =0$ and massless gauge and matter fields, 
and the broken-symmetric phase with $\langle\phi\rangle = \phi_+(T)$, with
massive gauge and matter fields.
The loop contributions from the gauge fields generate a cubic term
in the effective potential, which creates a barrier separating the two phases.
This result depends on a perturbative evaluation of the effective potential,
which presents problems for large Higgs masses as I will discuss later.
At 1-loop, the potential can be written as \cite{AH}
\begin{equation}
\label{eq:VEW}
V_{{\rm EW}}(\phi,T)=D\left (T^2-T_2^2 \right )\phi^2-ET\phi^3+{1\over
4}\lambda_T\phi^4,
\end{equation}
where the constants $D$ and $E$ are given by 
$$D=\left
[6(M_W/\sigma)^2+3(M_Z/\sigma)^2+6(M_T/\sigma)^2\right ]/24~\sim~0.17~,$$
and
$$E=\left
[6(M_W/\sigma)^3+3(M_Z/\sigma)^3\right ]/12\pi~\sim ~ 0.01~,$$
where I used,  $M_W=80.6$ GeV, $M_Z=91.2$ GeV, $M_T=174$ GeV \cite{TOP},
and $\sigma=246$ GeV. The (lengthy) expression for
$\lambda_T$, the temperature corrected Higgs self-coupling, can be found
in Ref. \cite{AH}. 
Here $T_2$ is the
temperature at which the origin becomes an inflection point ({\it i.e.},
below $T_2$ the symmetric phase is unstable),
given by
$T_2=\sqrt{(M_H^2-8B\sigma^2)/4D}$ ,
where the physical Higgs mass is given in terms of the 1-loop
corrected $\lambda$ as $M_H^2=\left (2\lambda+12B\right )
\sigma^2$, with $B=\left (6M_W^4+3M_Z^4-12M_T^4\right )/64\pi^2\sigma^4$.
For high temperatures, the system will
be in the symmetric phase with the potential exhibiting only one
minimum at $\langle \phi\rangle=0$. As the Universe expands and cools,
an inflection
point will develop away from the origin at
$\phi_{{\rm inf}}=3ET_1/2\lambda_T$, where $T_1=T_2/\sqrt{1-9E^2/8\lambda_TD}$.
For $T<T_1$, the inflection point separates into a
local maximum at $\phi_-(T)$ and a local minimum at $\phi_+(T)$,
with $\phi_\pm (T)=\{ 3ET\pm
\left [9E^2T^2-8\lambda_TD(T^2-T_2^2)\right ]^{1/2}\}/2\lambda_T$. At the
critical temperature, $T_C =T_2/\sqrt{1-E^2/\lambda_TD}$, 
the minima have the same free energy, $V_{{\rm EW}}(\phi_+,T_C)=V_{{\rm
EW}}(0,T_C)$. As $E\rightarrow 0$, $T_C\rightarrow T_2$ and the transition
is second order. Since $E$ and $D$ 
are fixed, the strength of the transition
is controlled by the value of the Higgs mass, or $\lambda$. 

Assuming that the above potential (or something close to it) correctly
describes the two phases, as the Universe cools belows $T_C$ the symmetric
phase becomes metastable and will decay by nucleation
of bubbles of the broken-symmetric phase which will grow and percolate
completing the transition. Departure from equilibrium will occur in the 
expanding bubble walls. This scenario relies on the assumption that the
transition is strong enough so that the usual homogeneous nucleation
mechanism correctly describes the approach to equilibrium. As I will discuss
later, this may not be the case for ``weak'' transitions. For now, we 
forget this problem and move on to briefly examine how to
generate the baryonic asymmetry with expanding
bubbles.

The last condition for generating baryon number is C and CP violation. It is
known that C and CP violation are present in the standard model. However,
the CP violation from the Kobayashi-Maskawa (KM) phase is too small to generate
the required baryon asymmetry. This is because the KM phase is multiplied
by a function of small couplings and  mixing angles, which strongly
suppresses the net CP
violation to numbers of order $10^{-20}$ \cite{CPVIOL}, 
while successful baryogenesis
requires CP violation of the order of $10^{-8}$ or so. A dynamical mechanism
to enhance the net CP violation in the standard model
was developed in detail by Farrar and 
Shaposhnikov \cite{FS}. It is based  on a phase separation of baryons via 
the scattering of quarks by the expanding bubble wall. This scenario
has been criticized by the authors of Ref. \cite{HUET} who claim that 
QCD damping effects reduce the asymmetry to a negligible amount. Even though
the debate is still going on, efficient baryogenesis within the standard
model is a remote possibility. 

For many, this is enough motivation to go
beyond the standard model in search of extensions which have an enhanced CP
violation built in. Several models have been proposed so far, although the
simplest invoke either more generations of massive fermions, or multiple
massive Higgs  doublets with additional CP violation in this sector of the
theory. Instead of looking into all models in detail, I will just briefly
describe the essential ingredients common to most models. 

The transition
is assumed to proceed by bubble nucleation. (For alternative mechanisms
based on topological defects, see Ref. \cite{Branden}.)
Outside the bubbles the Universe
is in the symmetric phase, and baryon number violation is occurring at the
rate $\Gamma \sim (\alpha_{{\rm W}}T)^4$. Inside the bubble the Universe is
in the broken symmetric phase and the rate of  baryon number violation is
$\Gamma\sim {\rm exp}[-\beta E_{{\rm S}}]$. Since we want any net excess
baryon number to be preserved in the broken phase, we must shut off the 
sphaleron rate inside the bubble. This imposes a constraint on the strength
of the phase transition, as $ E_{{\rm S}}\simeq \langle\phi (T)\rangle /g$; that
is, we must have a large ``jump'' in the vacuum expectation
value of $\phi$ during the transition, $\langle\phi (T)\rangle /T \gtrsim 1$, 
as
shown by Shaposhnikov 
\cite{CPVIOL}. 

Inside the bubble wall the fields are far from equilibrium
and there is CP violation, and thus a net asymmetry can be induced by the 
moving wall. In practice, computations are complicated by several factors,
such as the dependence on the net asymmetry on the bubble velocity and
on its thickness \cite{MCLERRAN}. Different charge transport
mechanisms based on leptons as opposed to quarks have been proposed, which
enhance the net baryonic asymmetry produced \cite{TUROK}. However, the basic
picture is that as matter traverses the moving wall an asymmetry is produced.
And since baryon number violation is suppressed inside the bubble, a net
asymmetry survives in the broken phase.  Even though no compelling model
exists at present, and several open questions related to the complicated
nonequilibrium dynamics remain, it is fair to say that the correct
baryon asymmetry may have been generated during the electroweak phase 
transition, possibly in some extension of the standard model. However,
I would like to stress that this conclusion
has two crucial assumptions built in it; that we know how to compute the
effective potential reliably, and that the transition is strong enough to
proceed by bubble nucleation.
In the next Section I
briefly discuss some of the issues involved and how they may be concealing
interesting new physics.

\section{ Challenges to Electroweak Baryogenesis}             

\subsection{The Effective Potential}
                               
A crucial ingredient in the computation of the net baryon number generated
during the electroweak phase transition is the effective potential. In order
to trust our predictions, we must be able to compute it reliably. However,
it is well known that perturbation theory is bound to fail due to severe
infrared problems. It is easy to see why this happens. At finite temperatures,
the loop expansion parameter involving gauge fields is $g^2T/M_{{\rm gauge}}$.
Since $M_{{\rm gauge}}=g\langle\phi\rangle$, in the neighborhood of $\langle
\phi\rangle=0$ the expansion diverges. This behavior can be improved by 
summing over ring, or daisy, diagrams \cite{BH}. 
However, the validity of the ring-improved effective potential
for the temperatures of interest relies on cutting off higher-order
contributions by invoking a nonperturbative magnetic plasma mass,
$M_{\rm plasma}$, for the gauge bosons such that the loop expansion
parameter, $g^2T/M_{\rm plasma}$, is less than 1. Since this
nonperturbative contribution is not well understood at present, one
should take the results from the ring-improved potentials with some
caution. Recent estimates show that perturbation theory breaks
down for Higgs masses above $70$ GeV \cite{BUCH}. These estimates are confirmed
by an alternative nonperturbative approach
based on the subcritical bubbles method \cite{GR}. 

Another problem that  appears in the evaluation of the effective potential
is due to loop corrections involving the Higgs boson. For second order phase
transitions, the vanishing of the effective potential's
curvature at the critical temperature leads
to the existence of critical phenomena characterized by diverging correlation
lengths. Even though there is no infrared-stable fixed point for first order
transitions, for large Higgs masses the transition is weak enough to induce
large fluctuations about equilibrium; the mean-field estimate for the 
correlation length $\xi(T)=M^{-1}(T)$ is certainly innacurate. 
The loop expansion parameter of the
effective static 3d theory is $\lambda T/M_{{\rm H}}(T)$, which diverges
as $T_{\rm C}\rightarrow T_2$ \cite{GK}. This behavior has led some authors
\cite{GK,AY} to invoke $\varepsilon$-expansion methods to deal with
the
infrared divergences. Although this is a promising line of work, it relies
on the success these methods have on different systems.
Another alternative is to go
to the computer and study the equilibrium properties of the
standard model on the lattice \cite{EWLATTICE}.
Recent results are encouraging inasmuch as they seem to be consistent with
perturbative results in the broken phase
for fairly small Higgs masses. Furthermore, they indicate how the transition
becomes weaker for large values of the Higgs mass, $M_{\rm H}\gtrsim 60$ GeV; 
physical quantities 
characterizing the strength of the transition, such as
the bubble's surface tension and the released latent heat, turn out to be quite
small. 
Let me move on to
discuss nonequilibrium aspects of the transition.

\subsection{Weak vs. Strong First Order Transitions}

In order to avoid the erasure of the produced net baryon number inside
the broken-symmetric phase,
the sphaleron rate must be suppressed within the
bubble. As mentioned earlier,
this amounts to imposing a large enough ``jump'' on the vacuum expectation value
of $\phi$ during the transition. In other words, the transition cannot be too
weakly first order. But what does it mean, really, to be ``weakly'' or 
``strongly'' first order? Looking into the literature, the most obvious
distinction between weak and strong is the thickness of the bubble. A 
``strong''
transition has thin-wall bubbles, that is, the bubble wall is much thinner
than the bubble radius (hence the name ``bubble''), while ``weak'' transitions
have thicker walls. In general, it is implicitly 
assumed that weak transitions proceed by the
usual bubble nucleation mechanism which, nevertheless, is derived only
for the case of strong
transitions. 

This is a very important point which must not be overlooked (although it
often is!);
the vacuum decay formalism used for the computation of nucleation rates relies
on a semi-classical expansion of the effective action. That is, we assume
we start at a {\it homogeneous} phase of false vacuum, and evaluate the
rate by summing over small amplitude fluctuations about the metastable
state \cite{LANGER}. This approximation must break down for weak
enough transitions, when we expect large fluctuations to be present within
the metastable phase. An explicit example of this breakdown was
recently discussed, where the extra free energy
available due to the presence of large-amplitude fluctuations was
incorporated into the computation of the decay rate \cite{GH}.

In Ref. \cite{GKII}, it was suggested that weak transitions may evolve by
a different mechanism, characterized by substantial mixing of the two phases
as the critical temperature is approached from above ({\it i.e.} as the 
Universe cools to $T_{{\rm C}}$). They estimated the fraction of the
total volume occupied by the broken-symmetric phase by assuming that the
dominant fluctuations about equilibrium  are subcritical bubbles of roughly
a correlation volume which interpolate between the two phases. Their approach
was later refined by the authors of Ref. \cite{GG} who found, within their
approximations, that the 1-loop electroweak potential shows considerable
mixing for $M_{{\rm H}}\gtrsim 55$ GeV. Clearly, the presence of 
large-amplitude, nonperturbative thermal fluctuations compromises the validity
of the effective potential, since it does not incorporate such corrections.

In order to understand the shortcomings of the mean-field approximation
in this context,
numerical simulations in 2d \cite{MG} and 3d \cite{BG} were performed, which
focused on the amount of ``phase mixing'' promoted by thermal fluctuations.
The idea was to simulate the nonequilibrium dynamics of a self-interacting
real scalar field, which is coupled to a thermal bath at temperature $T$. In
order to study the approximate behavior relevant to the electroweak phase
transition, the field was chosen to have a potential given by Eq. 1. (Note that
the temperature dependence of the potential can be scaled away with a 
proper redefinition of the couplings.) 
The coupling to the bath was modelled by a Markovian Langevin equation, which
assumes that the bath thermalizes much faster than any relevant dynamical
time-scale for the scalar field. Thus, the equation represents a coarse-grained
description of the dynamics, with faster modes with $\lambda << \xi(T)$
integrated out, where $\xi(T)=m^{-1}(T)$ is the mean field correlation length. 

The results show that the problem
boils down to how well localized the system is about the symmetric phase as it
approaches the critical temperature. If the system is well localized about the
symmetric phase, it will become metastable as the temperature drops below 
$T_{{\rm C}}$ and the transition can be called ``strong''. In this case,
the mean-field approximation is reliable. Otherwise, 
large-amplitude fluctuations away from the symmetric phase
rapidly grow, causing 
substantial mixing between the two phases. This will be a ``weak'' transition,
which will not evolve by bubble nucleation. 
Defining $\langle\phi\rangle_{{\rm V}}$
as the volume averaged field and $\phi_{{\rm inf}}$ as the 
inflection point nearest to the $\phi=0$ minimum, the criterion for a strong 
transition can be written as \cite{MG}
\begin{equation}
\langle\phi\rangle_{{\rm V}} < \phi_{{\rm inf}} ~~\: .
\end{equation}

Recently, an analytical model, based on the subcritical bubbles method,
was shown to qualitatively and {\it quantitatively} describe the results
obtained by the 3d simulation \cite{GHK}. The fact that subcritical
bubbles successfully model the effects of thermal fluctuations promoting
phase mixing and the breakdown of the mean-field approximation with
subsequent symmetry restoration, supports
previous estimates which showed that the assumption of homogeneous nucleation
is incompatible with standard model baryogenesis for $M_H\lesssim 55$ 
GeV \cite{GG,GR}. It is straightforward to adapt these computations to
extensions of the standard model. Thus, the requirement that the transition
proceeds by bubble nucleation can be used, together with the 
subcritical bubbles method, to constrain
the parameters 
of the potential.

In conclusion, the past few years saw encouraging progress towards the goal
of computing the baryon asymmetry of the Universe. Likewise, it 
has also become clear
that serious challenges lie ahead if we are to finally achieve this goal. The
need for enhanced CP violation probably calls for physics beyond the standard
model. Although this is an exciting prospect for many, we need guidance
from experiments in order to point us in the right direction. We must also
be able to compute
the effective potential reliably for a wider range of Higgs masses.
And finally, we must understand several nonequilibrium aspects of the phase
transition, be it within the context of expanding critical bubbles for strong
transitions or the dynamics of phase separation
for weak transitions. Judging from what has happened in the past few years,
progress will keep coming fast, and the goal will keep getting closer. 

I am grateful to my collaborators Rocky Kolb, Andrew Heckler,
Graciela Gelmini, 
Mark Alford, Julian Borrill,
and Rudnei Ramos for the many long discussions on bubbles and
phase transitions. I am also grateful to Fernando Brandt and the local
organizing committee for their warm hospitality during this Conference.
This work was partially supported by the 
National Science Foundation
through a Presidential Faculty Fellows
Award no. PHY-9453431 and by a NASA grant no. NAGW-4270.


\begin{thebibliography}{99}


\bibitem{KT} For reviews see, L. Yaffe, hep-ph/9512265, 
A. G. Cohen, D. B. Kaplan, and A. E. Nelson, Annu. Rev. Nucl. Part.
Sci. {\bf 43}, 27 (1993); A. Dolgov, Phys. Rep. {\bf 222}, 311 (1992); 
K. A. Olive, 
in
``Matter under extreme conditions'', eds. H. Latal and W. Schweiger 
(Springer-Verlag, Berlin, 1994).

\bibitem{STEIGMAN} G. Steigman, Ann. Rev. Astron. Astrophys. {\bf 14}, 339 
(1976).

\bibitem{SAKHAROV} A. D. Sakharov, JETP Lett. {\bf 5}, 24 (1967).

\bibitem{GUTBAR} For a review see E. W. Kolb and M. S. Turner, Ann. Rev.
Nucl. Part. Sci. {\bf 33}, 645 (1983); {\it ibid.} The Early Universe,
(Addison-Wesley, Redwood, CA, 1990).

\bibitem{REHEAT} L. Kofman, A. Linde, and A.A. Starobisnky, Phys. Rev.
Lett. {\bf 73}, 3195 (1994); D. Boyanosvsky {\it et al}, hep-ph/9511361,
hep-ph/9507414; hep-ph/9505220; D. Kaiser, hep-ph/9507108.

\bibitem{tHOOFT} G. t'Hooft, Phys. Rev. Lett. {\bf 37}, 8 (1976); Phys. Rev.
D{\bf 14}, 3432 (1976).

\bibitem{KRS} V. A. Kuzmin, V. A. Rubakov,
and M. E. Shaposhnikov, Phys. Lett. {\bf B155}, 36 (1985).


\bibitem{SPHALERON} N. S. Manton, Phys. Rev. D{\bf 28}, 2019 (1983);
F. R. Klinkhammer and N. S. Manton, Phys. Rev. D{\bf 30}, 2212 (1984).

\bibitem{HIGHTSPH} P. Arnold and L. McLerran, Phys. Rev. D{\bf 36}, 581
(1987); {\it ibid.} D{\bf 37}, 1020 (1988); J. Ambjorn and A. Krasnitz,
Phys. Lett. {\bf B362}, 97 (1995).

\bibitem{AH}  G. W. Anderson and L. J. Hall, Phys. Rev. D
{\bf 45},
2685 (1992); M. Dine, P. Huet, and R. Singleton, Nucl. Phys.
{\bf B375}, 625 (1992).

\bibitem{TOP} F. Abe et al. (CDF Collaboration), Phys. Rev. Lett. {\bf 73},
225 (1994).

\bibitem{Branden} R. Brandenberger and A.-C. Davis, Phys. Lett. {\bf B308},
79 (1993); R. Brandenberger,  A.-C. Davis, and M. Trodden, Phys. Lett. {\bf 
B335}, 123 (1994).

\bibitem{CPVIOL} C. Jarlskog, Z. Physik C{\bf 29}, 491 (1985); Phys. Rev.
Lett. {\bf 55}, 1039 (1985); M. E. Shaposhnikov, Nucl. Phys. B{\bf 287},
757 (1987); B{\bf 299}, 797 (1988).

\bibitem{FS} G. R. Farrar and M. E. Shaposhnikov, Phys. Rev. Lett.
{\bf 70}, 2833 (1993); Phys. Rev. {\bf D50}, 774 (1994).

\bibitem{HUET}  M. B. Gavela,
P. Hern\'andez, J. Orloff, O. P\`ene, and C. Quimbay, Nucl. Phys. {\bf B430},
382 (1994)
P. Huet and E. Sather, preprint SLAC-PUB-6479 (1994).

\bibitem{MCLERRAN}  B. Liu, L. McLerran, and N. Turok, Phys. Rev.
{\bf D46}, 2668 (1992); M. Dine and S. Thomas, Phys. Lett. {\bf B328}, 73 
(1994); G.D. Moore and T. Prokopec, Phys. Rev. Lett. {\bf 75}, 777 (1995).

\bibitem{TUROK} A. G. Cohen, D. B. Kaplan, and A. E. Nelson, Phys. Lett.
B{\bf 245}, 561 (1990); Nucl. Phys. B{\bf 349}, 727 (1991); B{\bf 373}, 
453 (1992); Phys. Lett. {\bf B336}, 41 (1994); 
M. Joyce, T. Prokopec, and N. Turok, Phys. Rev. Lett. {\bf 75}, 1695 (1995);
hep-ph/9410281.

\bibitem{BH}  P. Arnold and O. Espinosa, Phys. Rev. {\bf D47}, 3546
(1993); M. Dine, P. Huet, R.G. Leigh, A. Linde, and D. Linde, Phys. Rev.
{\bf D46}, 550 (1992); C.G. Boyd, D.E. Brahm, and S. Hsu, Phys. Rev.
{\bf D48}, 4963 (1993); M. Quiros, J.R. Spinosa, and F. Zwirner, Phys. Lett.
{\bf B314}, 206 (1993); W. Buchm\"uller, T. Helbig, and D. Walliser,
Nucl. Phys. {\bf B407}, 387 (1993); M. Carrington, Phys. Rev. {\bf D45},
2933 (1992).


\bibitem{BUCH} W. Buchm\"uller and Z. Fodor, Phys. Lett. B{\bf 331}, 124
(1994).

\bibitem{GR} M. Gleiser and R. Ramos, Phys. Lett. B{\bf 300}, 271 (1993).

\bibitem{GK} M. Gleiser and E.W. Kolb,  Phys. Rev. {\bf D48}, 1560 (1993). 

\bibitem{AY}  P. Arnold and L. Yaffe, Phys. Rev. {\bf D49},
3003 (1994).

\bibitem{EWLATTICE}  K. Farakos, K. Kajantie, K. Rummukainen, and M.
Shaposhnikov, hep-lat/9510020; Nucl. Phys. {\bf B407}, 356 (1993); {\it ibid.}
{\bf B425}, 67 (1994); {\it ibid.}{\bf B442}, 317 (1995);
B. Bunk, E.-M. Ilgenfritz, J. Kripfganz,
and A. Schiller, Phys. Lett. {\bf B284}, 372 (1992); Nucl. Phys. {\bf B403},
453 (1993); Z. Fodor et al. Nucl. Phys. {\bf B439}, 147 (1995).

\bibitem{LANGER}  J. S. Langer,
Ann. Phys. (NY) {\bf 41}, 108 (1967);
{\it ibid.} {\bf 54}, 258 (1969);
  M. B. Voloshin, I. Yu. Kobzarev, and L. B. Okun',
        Yad. Fiz. {\bf 20}, 1229 (1974)
        [Sov. J. Nucl. Phys. {\bf 20}, 644 (1975);
S. Coleman, Phys. Rev. {\bf D15}, 2929 (1977); C. Callan
and S. Coleman, Phys. Rev. {\bf D16}, 1762 (1977);  A. D. Linde, 
Nucl. Phys. {\bf B216}, 421 (1983);
[Erratum:
{\bf B223}, 544 (1983)]; M. Gleiser, G. Marques, and R. Ramos, Phys. Rev. 
{\bf D48}, 1571 (1993); D. Brahm and C. Lee, Phys. Rev. {\bf D49}, 4094 (1994).

\bibitem{GH} M. Gleiser and A. Heckler, Phys. Rev. Lett. {\bf 76}, 180 (1996).

\bibitem{GKII}  M. Gleiser and E. W. Kolb, Phys. Rev. Lett. {\bf 69},
1304 (1992);  M. Gleiser, E. W. Kolb, and
R. Watkins, Nucl. Phys. {\bf B364}, 411 (1991); N. Tetradis, Z. Phys.
{\bf C57}, 331 (1993).


\bibitem{GG} G. Gelmini and M. Gleiser, Nucl. Phys. B{\bf 419}, 129 (1994).

\bibitem{MG} M. Gleiser,
Phys. Rev. Lett. {\bf 73}, 3495 (1994).

\bibitem{BG} J. Borrill and M. Gleiser, Phys. Rev. {\bf D51}, 4111 (1995).

\bibitem{GHK} M. Gleiser, A. Heckler, and E.W. Kolb, cond-mat/9512032, 
submitted to Physical Review Letters.


\end{thebibliography}
\end{document}